\documentclass[conference]{IEEEtran}
\ifCLASSINFOpdf
\else
\fi

\usepackage{amssymb}

\usepackage[cmex10]{amsmath}
\usepackage{graphicx}
\usepackage{caption}
\usepackage{subfigure}
\usepackage{multicol}
\usepackage{soul}

\usepackage{lipsum}
\usepackage{adjustbox}
\usepackage[figuresright]{rotating}

\begin{document}
%
\title{Evaluating Impact of Human Errors on the Availability of Data Storage Systems}



%

\author{\IEEEauthorblockN{Mostafa Kishani,
Reza Eftekhari,
and Hossein Asadi}
\IEEEauthorblockA{Data Storage, Networks, \& Processing (DSN) Lab, Department of Computer Engineering, Sharif University of Technology}}


\maketitle
\vspace{-0.5cm}
\begin{abstract}
In this paper, we investigate the effect of incorrect disk replacement service on the availability of data storage systems.
To this end, we first conduct Monte Carlo simulations to evaluate the availability of disk subsystem 
by considering disk failures and incorrect disk replacement service.
We also propose a Markov model that corroborates the Monte Carlo simulation results.
We further extend the proposed model to consider the effect of automatic disk fail-over policy. 
The results obtained by the proposed model show that overlooking the impact of incorrect disk replacement can result up to three orders of magnitude unavailability underestimation. Moreover, this study suggests that by considering the effect of human errors, the conventional believes about the dependability 
of different RAID mechanisms should be revised. 
The results show that in the presence of human errors, RAID1 can result in 
lower availability compared to RAID5.
\end{abstract}


%
\IEEEpeerreviewmaketitle

\vspace{-0.3cm}
\section{Introduction}
\label{sec:Intro}
Human errors have significant impact on the availability of Information systems
~\cite{oppenheimer2003internet,brown2001err,oppenheimer2003importance}
 where some field studies have reported that 19\% of system failures are
caused by human errors~\cite{haubert,oppenheimer2003importance}. 
In large data-centers with \emph{Exa-Byte} (EB) storage capacity (by employing more than one million disk drives), one should expect at least a disk failure per hour. 
Despite using mechanisms such as automatic fail-over in modern data-centers, in many cases the role of human agents is inevitable.
Meantime, the probability of human error, even by using precautionary mechanisms such as checklists and employing high-educated and high-trained
human resources, is between 0.1 and 0.001 ~\cite{NASA-her,gibson2006feasibility,us1975reactor,swain1983handbook}.
Such statistics translate that an exascale data-center will face multiple human errors a day. 

Disk drives are of most vulnerable components in a \emph{Data Storage System} (DSS). 
Disk failures and \emph{Latent Sector Errors} (LSEs) ~\cite{Schroeder-2010-TOS} 
are of main sources of data loss in a disk subsystem.
Several studies have tried to investigate the effect of these two incidences on a single disk and disk array reliability 
~\cite{Schroeder-2010-TOS,Elerath-2009-TC,Greenan-HOTSTORAGE-2010,Schroeder-FAST-2007}. 
In particular, the failure root cause breakdown in previous studies~\cite{haubert,oppenheimer2003importance} shows that human error is of great importance.


In this paper, we propose an availability model for the disk subsystem of a \emph{Backed-up} data storage system\footnote{An storage system that keeps an 
updated backup of data, for example on a tape. In such system, we assume that data loss can be recovered using the backup and has just an unavailability consequence.} by considering 
the effect of disk failures and human errors.
While the incorrect repair service can have many different roots and happen in many different conditions, in this work we 
just consider the incorrect disk replacement service and call it \emph{Wrong Disk Replacement}.
In our analysis, both disk subsystems with and without automatic disk fail-over are considered. 
The proposed analytical technique is based on Markov models and hence requires the assumption of exponential distributions for \emph{time-to-failure} and \emph{time-to-restore}.
Furthermore, to cope with other probability distribution functions such as \emph{Weibull}, 
that describes the disk failure behavior in a more realistic manner~\cite{Schroeder-FAST-2007}, 
we have developed a model based on \emph{Monte-Carlo} (MC) simulations. 
This model has also been used as a reference to validate the proposed Markov model
when using exponential distributions.

By incorporating the impact of human errors on the availability of disk subsystem, 
several important observations are obtained. 
First, it is shown that overlooking the impact of incorrect repair service will result in 
a considerable underestimation (up to 263X) of the system downtime. 
Second, it is observed that in the presence of human errors,  
conventional assumptions on the availability ranking of different \emph{Redundant Array of Independent Disks} (RAID) configurations 
can be contradicted.  
Third, it is demonstrated that automatic disk fail-over can significantly improve the overall system availability 
when on-line rebuild is provided by using spare disks. 

The remainder of this paper is organized as follows.
Section~\ref{sec:related} represents a background on human errors. 
Section~\ref{sec:ref-avail} elaborates the Monte-Carlo simulation-based model. 
Section~\ref{sec:proposed} presents the proposed Markov models, considering the impact of human errors.
Section~\ref{sec:results} provides simulation results 
and the corresponding findings. 
Lastly, Section \ref{sec:Conclude} concludes the paper.

\vspace{-0.3cm}
\section{Background}
\label{sec:related}

\subsection{Human Error in Safety-Critical Applications}
\label{sec:human_error}

To better understand and quantify human errors in a non-benign system, the  \emph{Human Reliability Assessment} (HRA) ~\cite{swain1990human} techniques have been developed where its major focus is the quantification of \emph{Human Error Probability} ($hep$) which is simply defined by 
the fraction of error cases observed, over the opportunities for human errors~\cite{gibson2006feasibility}.
By collecting $hep$ values obtained by \emph{NASA}, \emph{EUROCONTROL}, and \emph{NUREG}, we found that human error has usually a probability in the range of $0.001$ to $0.1$ depending on the application and situation.
This probability mainly varies from $0.001$ up to $0.01$ in enterprise and safety-critical applications 
\cite{gibson2006feasibility,us1975reactor,swain1983handbook,NASA-her}. 

\subsection{Human Errors in Data Storage Systems}
While human errors in data storage systems can happen in very different situations, in this work we focus on one of the most 
prevalent samples, the wrong disk replacement.
In a RAID array, given RAID5, with no disk spare, the disk fail-over process can start after replacing the failed disk with a brand-new disk. 
Consider the case that the operator replaces the brand-new disk with one of the operating disks, rather than the failed one.
In this case, two disks are inaccessible (the failed disk and the operating, wrongly removed disk), making the entire data unavailable. 
However, detecting the human error and undoing the wrong disk replacement makes the array available at no data loss cost. 

In the next section, we describe a simulation-based reference model to evaluate the availability of data storage systems considering disk failures and the effect of human errors happening in disk replacement process.

\vspace{-0.1cm}
\section{Availability of a Backed-up Disk Subsystem by Monte-Carlo Simulation}
\label{sec:ref-avail}



\begin{figure}
\begin{centering}
\includegraphics[width=0.5\textwidth]{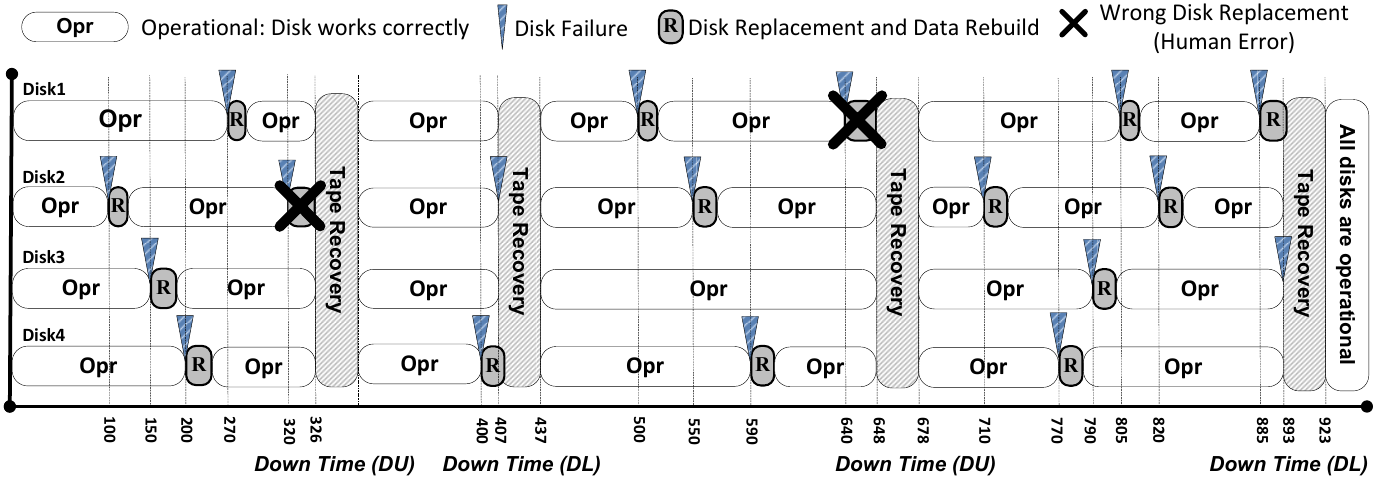}
\caption{MC Simulation to Calculate Availability of a RAID5 3+1 Array in Presence of Human Errors  (rebuild time = 10h)}
\vspace{-0.3cm}
\label{fig:raid5-simulation-events}
\par\end{centering}
\end{figure}

In the MC model, the failure and repair events are generated by assuming the desired distributions such as \emph{Weibull} and exponential. 
Fig.~\ref{fig:raid5-simulation-events} illustrates an example of the MC simulation for a RAID5 $(3+1)$ array. 
In case two consecutive disk failures happen in the same array 
and the second failure is before the recovery of the first failure, a data loss event happens (as shown in Fig.~\ref{fig:raid5-simulation-events}). 
As we assume that the data storage system is backed-up, we consider the data loss event as data unavailability, which duration is the data loss recovery (tape recovery in our example) time.
In the case of single disk failure, the failed disk is replaced by a human agent. 
However, the occurrence of a human error in the disk replacement process makes another working disk unavailable, resulting in the unavailability of the 
entire data array (in the case of RAID5), which duration varies by the human error recovery time.  
The overall availability of the disk subsystem is calculated 
by dividing the total disk subsystem \emph{uptime} 
by the overall \emph{simulation time}.
The error of MC simulations is inversely proportional to the root square of the number of iterations and the \emph{t-student} coefficient for a target confidence level~\cite{lange1989robust}. 

\vspace{-0.1cm}
\section{Availability of a Backed-up Disk Subsystem Using Markov Model}
\label{sec:proposed}
\vspace{-0.1cm}
In this section, 
we propose a Markov Model for a backed-up disk subsystem availability that corroborates
the Monte Carlo reference model by assuming an exponential distribution for both failure and repair rates.
Finally, we extend the Markov model for a disk subsystem with automatic fail-over.

\vspace{-0.2cm}
\subsection{Markov Model of RAID5 in Presence of Human Errors \label{sec:proposed-raid5}}
\begin{figure}
\begin{centering}
\includegraphics[width=3.5in]{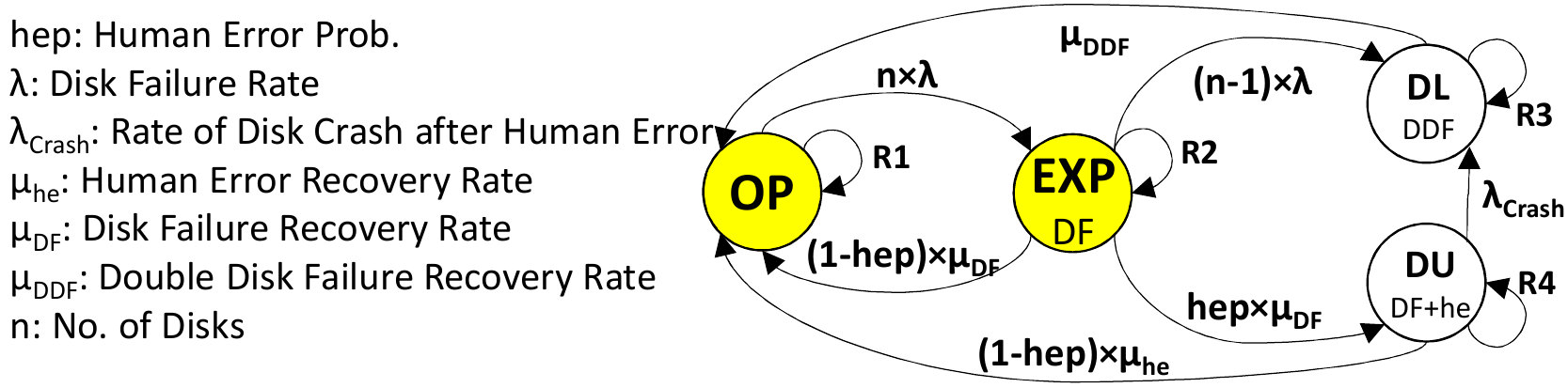}
\vspace{-0.6cm}
\caption{Markov Model for RAID5 Availability}
\vspace{-0.7cm}
\label{fig:raid5hr}
\par\end{centering}
\end{figure}

\vspace{-0.1cm}
Fig.~\ref{fig:raid5hr} shows the proposed Markov model for the availability of a backed-up disk subsystem by 
considering the effect of disk failures and human errors. 
In this model, disk failure rate, disk repair rate, double disk failure recovery rate from primary backup, and \emph{Human Error Probability} 
are shown by $\lambda$, $\mu_{DF}$, $\mu_{DDF}$, and $hep$, respectively. 
Upon the occurrence of the first disk failure, 
the system state will move from the operational ($OP$) to the exposed state ($EXP$).
While being in the exposed state, 
a second disk failure will lead to a \emph{Data Loss} (DL) event 
while a human error during disk replacement will lead to a \emph{Data Unavailability} (DU) event.
If the human agent successfully replaces the failed disk, the array returns to the $OP$ state.

When the array is in the $DU$ state, the incidence of human errors during the fail-over process makes the array to stay in the $DU$ state. 
Otherwise, if no human error happens in the fail-over process, the array state transits to the $OP$ state. 
In the $DU$ state, if the wrongly replaced disk is crashed, the array switches to $DL$.
Finally, when the array is in the $DL$ state, it can be recovered by the rate of $\mu_{DDF}$. 

\vspace{-0.2cm}
\subsection{Markov Model of RAID5 With Automatic Fail-over}
\label{sec:hot-spare-model}
\vspace{-0.2cm}
Here, we study the effect of automatic disk fail-over 
when on-line rebuild process is being performed using hot-spare disks. 
In the conventional disk replacement policy, a failed disk may be replaced by a new disk 
before the completion of the on-line rebuild process. 
In the automatic fail-over policy as opposed to the conventional disk replacement policy, 
the replacement process should be started after the completion of on-line rebuild process. 
In automatic fail-over policy, 
it is assumed that a hot spare disk is available within the array 
while the system is in the operational state.

\begin{figure}
\begin{centering}
\includegraphics[width=3.4in]{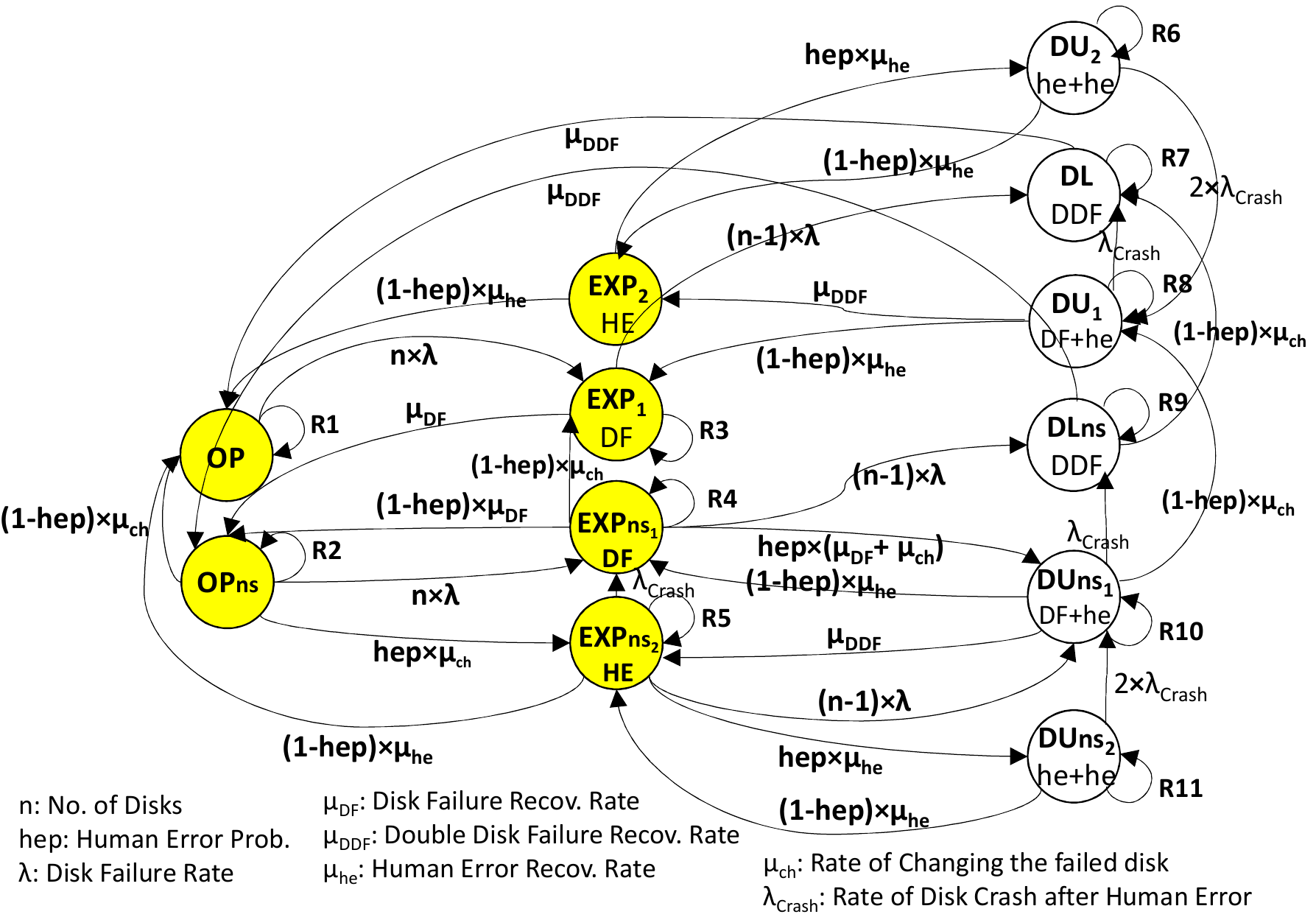}
\vspace{-0.2cm}
\caption{Markov Model of RAID5 Availability with Automatic Fail-over}
\vspace{-0.5cm}
\label{fig:raid5-spare}
\par\end{centering}
\end{figure}

Fig.~\ref{fig:raid5-spare} shows the Markov model of a RAID5 array employing the automatic fail-over policy.
The system is in the $OP$ state when all disks work properly and a spare disk is present. 
In the case of a disk failure, the array state switches to $EXP_1$. 
In the $EXP_1$ state, the system goes to either the $DL$ state by another disk failure or 
the $OP_{ns}$ state if the failed disk is rebuilt into the available spare disk.
In the $OP_{ns}$ state, all disks of the array work properly but no spare is present. 
Automatic fail-over paradigm forbids the operator 
to replace the affected disk before the completion of on-line rebuild process. 
Hence, disk replacement can be performed at the states other than the $EXP_1$ state and there is no possibility of human error in the $EXP_1$ state.
When the system is in the $OP_{ns}$ state, a disk failure switches the system state to $EXPns_1$. 
If the failed disk is successfully replaced by the new disk, the array returns to the $OP$ state.
Otherwise, if a human error happens in the disk replacement process, the array switches to $EXPns_2$. 

In the $EXPns_1$ state, the array has a failed disk and no spare.
In this state, successful disk fail-over process changes the array state to $OP_{ns}$.
Upon the successful replacement of failed disk in the $EXPns_1$ state, the array 
switches to $EXP_1$.
However, by happening a human error in the process of disk fail-over or in 
the process of failed disk replacement, 
the array switches to $DUns_1$.
Upon a disk failure when the array is in the $EXPns_1$ state, 
the array switches to $DLns$.

In the $EXPns_2$ state, one of the operational disks is wrongly replaced by the new disk due to human error. To remove this error,
the wrongly removed disk should be placed back and in return, the failed disk should be removed. 
If this process happens successfully, the array goes back to $OP$.
Otherwise, if another human error happens in the process of recovering the human error, the array switches to $DUns_2$.
In the $EXPns_2$ state, if the wrongly removed disk crashes,
the array switches to $EXPns_1$.
Happening a disk failure when the array is in the $EXPns_2$ state 
switches the array to $DUns_1$.

In the $DL$ state, the user data is totally lost due to a \emph{Double Disk Failure} (DDF) and 
a hot spare disk is available. 
In this case, DDF could be recovered by the rate of $\mu_{DDF}$.
Similarly in the $DLns$ state, the user data is totally lost due to a DDF but no spare is available.
Here, recovery from DDF changes 
the array state to $OP_{ns}$.
In the $DLns$ state, if one of the failed disks is successfully replaced by the new spare disk, the array switches to the $DL$ state.

In the $DUns_1$ state, the array is totally unavailable 
due to a disk failure and a human error.
In this state, the successful recovery of human error changes the 
array state to $EXPns_1$. 
However, if the wrongly removed disk crashes, the array switches to 
$DLns$.
In the $DUns_1$ state, performing the disk fail-over by using the disk array is not possible as the user data is unavailable due to the human error. In this case, performing the disk fail-over 
before recovering the human error is similar to the case of recovering DDF by the rate of $\mu_{DDF}$. 
In the $DUns_1$ state, if the failed disk is successfully replaced 
by the new spare disk, the array switches to the $DU_1$ state.

In the $DUns_2$ state, the array data is totally unavailable due to 
the occurrence of two human errors. 
In this case, the array can switch to $EXPns_2$ if one of the human errors is successfully recovered.
However, if one of the wrongly removed disks crashes, the array switches to $DUns_1$.

$EXP_2$ is similar to $EXPns_2$ except the point 
that the hot spare disk is available in this state.
Similarly, the $DU_1$ and $DU_2$ states are similar to  
$DUns_1$ and $DUns_2$, respectively, except the point that the hot spare disk is available in $DU_1$ and $DU_2$.


Comparing the Markov model of a RAID5 array employing the automatic fail-over (Fig.~\ref{fig:raid5-spare}) 
and a RAID5 array using the conventional disk replacement policy (Fig.~\ref{fig:raid5hr}) 
shows a longer path from $OP$ to $DU$ state when the automatic fail-over is performed in the system. 
Hence, it can be realized that the probability of being in the $DU$ state significantly 
decreases by using the automatic fail-over policy. 
Detailed numerical results of this model and comparison with a RAID array 
employing the conventional replacement policy will be presented in Section~\ref{sec:hotSpareReport}.

\vspace{-0.2cm}
\section{Experimental Setup and Simulation Results}
\label{sec:results}
\vspace{-0.1cm}

\subsection{Validation of Markov Model with Simulation-Based Model}
\label{sec:markovValidation}

\begin{figure}
\begin{centering}
\includegraphics[width=2.6in]{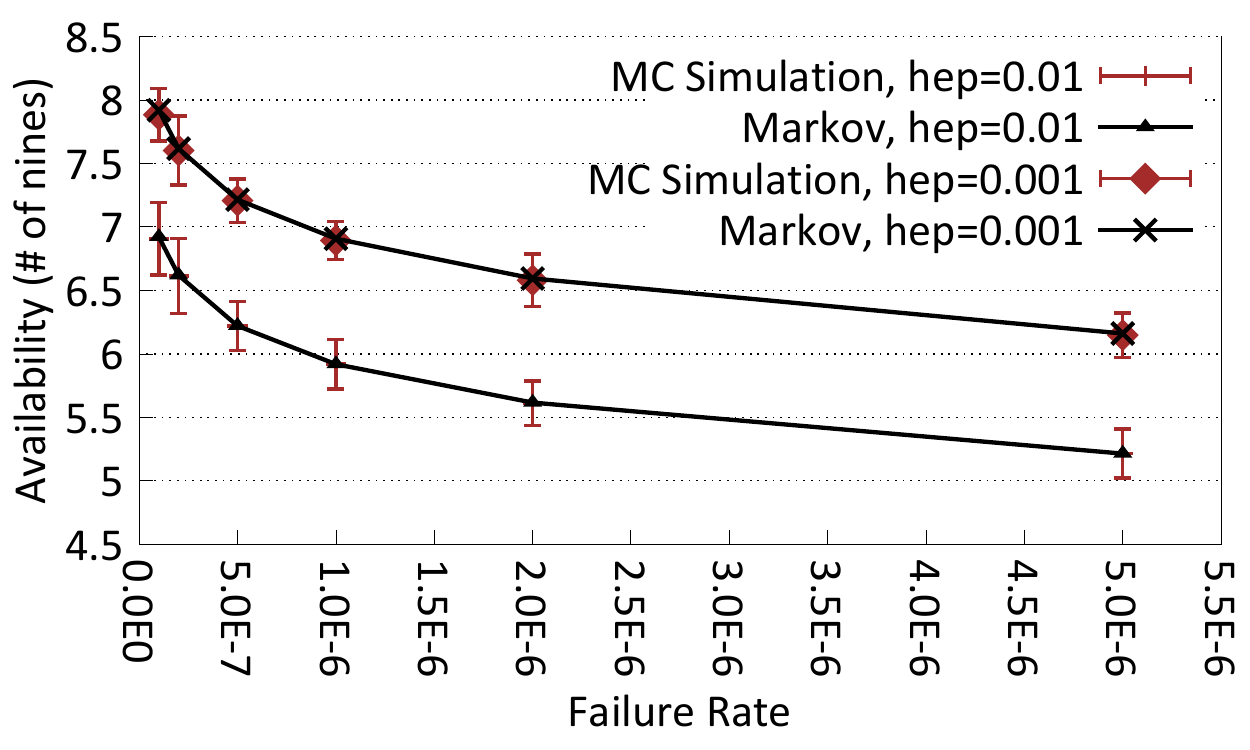}
\vspace{-0.3cm}
\caption{Comparison of Reference and Markov Models}
\vspace{-0.4cm}
\label{fig:mc-markov-comparison}
\par\end{centering}
\end{figure}

Fig.~\ref{fig:mc-markov-comparison} shows the comparison of the proposed MC simulation results (for $10^6$ iterations and 99\% confidence level) 
and the availability values obtained by the Markov model. 
As shown in this figure, the availability values obtained by the Markov model are
within the error interval of the results obtained by the MC simulations for both $hep=0.001$ and $hep=0.01$. 

\vspace{-0.2cm}
\subsection{Availability Estimation in Presence of Human Error}
\vspace{-0.1cm}

\begin{figure}
    \centering
    \hspace{-1cm}
  
        \includegraphics[width=0.26\textwidth]{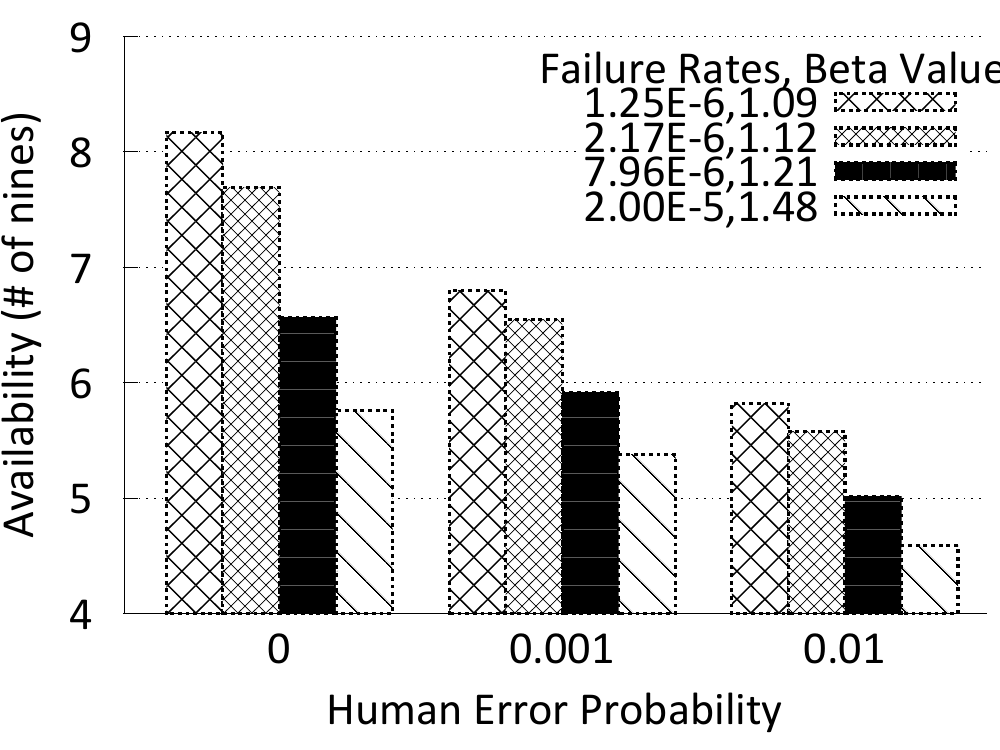}
     \vspace{-0.3cm}
    \caption{Availability of a RAID5(3+1) Array}
    \vspace{-0.6cm}
    \label{fig:lse-0}
\end{figure}

Fig.~\ref{fig:lse-0} reports the availability results of a RAID5 $3+1$ array 
in the presence of human errors for different disk failure rates.
The availability of the disk subsystem has been reported for the traditional availability model (assuming $hep$=$0$)
as well as two different human error probabilities ($hep=0.001$ and $hep=0.01$). 
We consider typical values for the repair rate in our experiments. In particular, we consider 0.1 and 0.03 values for $\mu_{DF}$ and $\mu_{DDF}$, respectively.
We also consider $\mu_{s}=1$, $\mu_{he}=1$, and $\lambda_{crash}=0.01$.
Considering a constant failure rate (for example, $\lambda=10^{-6}$),
it is observed that the availability of a disk subsystem 
is inversely proportional with human error probability.
As the results show, with the human error probability equal to 0.001, 
the availability of the disk subsystem drops between one to two orders of magnitude. 



\vspace{-0.2cm}
\subsection{Availability Comparison of RAID Configurations with Equivalent Usable Capacity}
\vspace{-0.15cm}

\begin{figure}
\begin{centering}
\includegraphics[width=1.5in]{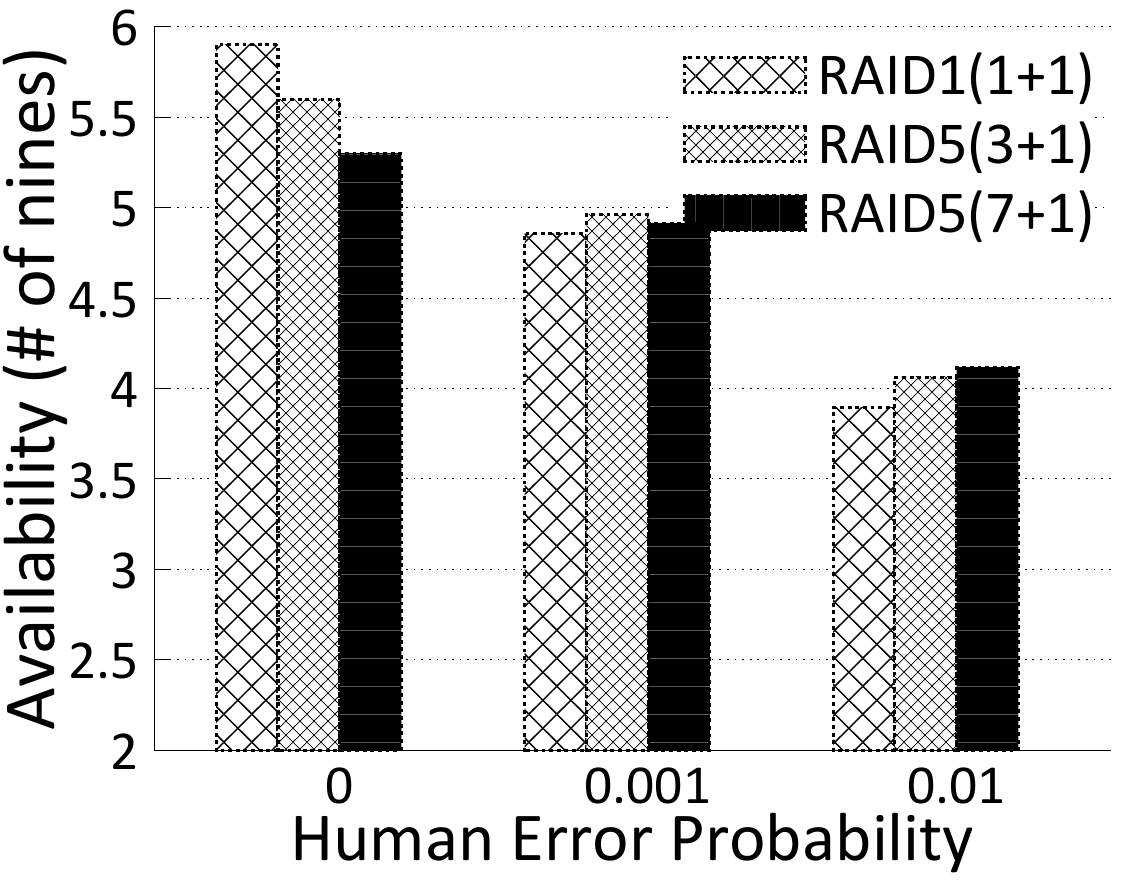}
\caption{Comparison of Availability of RAID Configurations with Equivalent Usable Capacity and Exponential Disk Failure}
\vspace{-0.5cm}
\label{fig:usable-exp}
\par\end{centering}
\end{figure}

Fig.~\ref{fig:usable-exp} compares the availability of three different RAID configurations including $R5(3+1)$, $R5(7+1)$, and $R1(1+1)$ 
with equivalent usable (logical) capacity,  
in the presence of human errors ($hep$=$0$, $hep$=$0.001$, and $hep$=$0.01$), assuming exponential failure distribution ($\lambda=10^{-5}$). 
Comparing the three RAID configurations by assuming no human errors ($hep=0$) shows that $RAID1 (1+1)$ results in a higher availability compared to 
$RAID5 (3+1)$ and $RAID5 (7+1)$. 
However, by considering $hep=0.001$, the availability of all RAID configurations dramatically decrease, while our results show a more significant decrease 
in the $RAID1 (1+1)$ configuration, making its availability slightly lower than both $RAID5 (3+1)$ and $RAID5 (7+1)$ configurations. 
This can be described by the higher \emph{Effective Replication Factor\footnote{The ratio of storage physical size to 
the logical (usable) size~\cite{muralidhar2014f4}.}} (ERF) of 
$RAID1 (1+1)$ ($ERF=2$) compared to $RAID5 (3+1)$ ($ERF=1.33$) and $RAID5 (7+1)$ ($ERF=1.14$), which mandates employing higher number of disks for a specific
usable capacity, increasing the chance of disk failure and consequently, human errors.
By considering higher $hep$ values (e.g., $0.01$), we observe more gap between RAID configurations where both $RAID1 (1+1)$ and $RAID5 (5+1)$ 
show lower availability compared to $RAID5 (7+1)$, that can again be described by the lower ERF of $RAID5 (7+1)$.
 
\vspace{-0.2cm}
\subsection{Effect of Automatic Disk Fail-over Policy}
\label{sec:hotSpareReport}
\vspace{-0.2cm}

\begin{figure}
    \centering
    \hspace{-1cm}
   
        \includegraphics[width=0.28\textwidth]{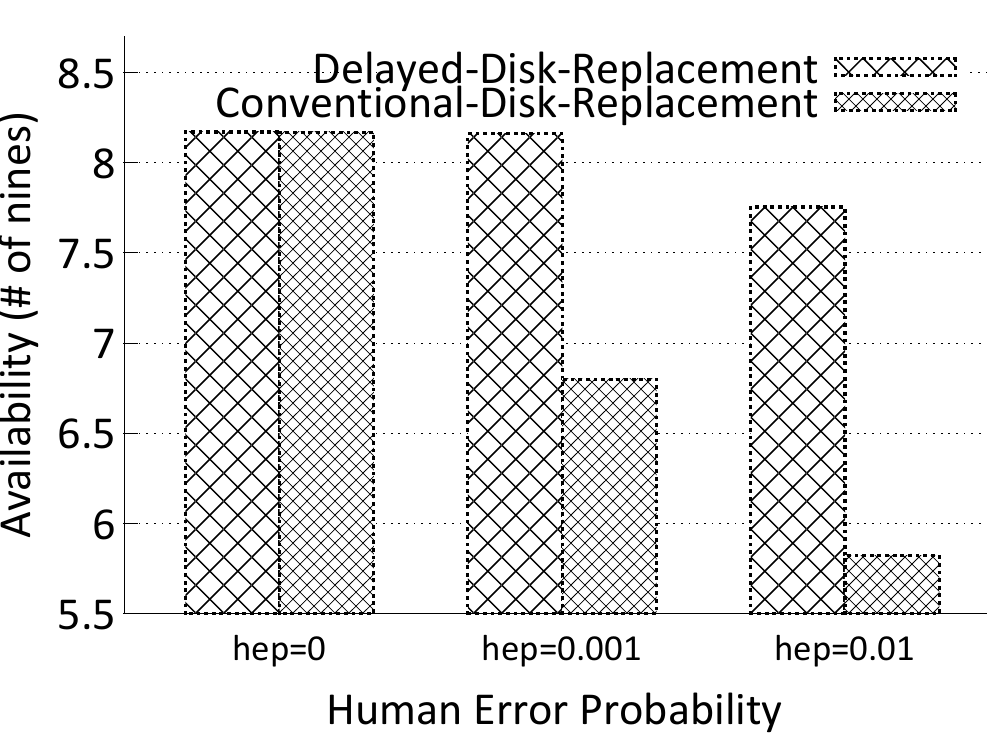}
       \vspace{-0.3cm}
    \caption{Availability of Automatic Fail-over Policy}
    \vspace{-0.7cm}
    \label{fig:hot-spare}
\end{figure}

In this section, we report the effect of the automatic fail-over policy  
when on-line rebuild process is being performed using hot-spare disks.
Fig.~\ref{fig:hot-spare} compares the availability of two RAID5 arrays, 
performing conventional and automatic fail-over in the presence of human errors. 
As the results show, using automatic fail-over policy can significantly moderate the effect of human errors. 
For example, assuming $hep=0.01$, automatic fail-over increases the system availability by two orders of magnitude 
as compared to the conventional disk replacement policy.  
The results reported in Fig.~\ref{fig:hot-spare} also demonstrate that
the delayed replacement policy shows higher availability improvement when $hep$ has greater values. 

\vspace{-0.2cm}
\section{Conclusion and Future Works}
\label{sec:Conclude}
\vspace{-0.2cm}
In this paper, we investigated the effect of incorrect disk replacement service on the availability of a backed-up
disk subsystem by using Monte Carlo simulations and Markov models.
By taking the effect of incorrect disk replacement service into account, 
it is shown that a small percentage of human errors (e.g., $hep=0.001$) 
can increase the system unavailability by more than one order of magnitude. 
Using the proposed models, 
it is also shown that in some cases the dependability ranking of RAID configurations is not as conventional. 
Additionally, it is shown that automatic fail-over can increase the system availability by two orders of magnitude. 
Such observations can be used by both designers and system administrators to enhance the overall system availability.
\vspace{-0.2cm}
\bibliographystyle{IEEEtran} 
\bibliography{mss-ras-bib}

\end{document}